\documentclass[preprint,showpacs,preprintnumbers,amsmath,amssymb]{revtex4}

\usepackage{graphicx}
\usepackage{dcolumn}
\usepackage{bm}

\usepackage{feynmf}
\usepackage{slashed}

\begin{document}

\title{Off-Shell Tachyons}
\author{Yi-Lei Tang}
\affiliation
{Institute of Theoretical Physics, Chinese Academy of Sciences, \\
and State Key Laboratory of Theoretical Physics,\\
P. O. Box 2735, Beijing 100190, China}
 \email{tangyilei10@itp.ac.cn}

\date{\today}

\begin{abstract}
The idea that the new particles invented in some models beyond the standard model can appear only inside the loops is attractive. In this paper, we fill these loops with off-shell tachyons, leading to a solution of the zero results of the loop diagrams involving the off-shell non-tachyonic particles. We also calculate the Passarino-Veltman $A_0^o$ and $B_0^o$ of the off-shell tachyons.
\end{abstract}


\maketitle
\section{Introduction}

Recently, an interesting and attractive idea that all the supersymmetric particles could only appear inside the loops has been introduced \cite{ClaimToBeEarlest, MainSource}. By modifying the quantization techniques of the supersymmetric particles, they cannot appear in the out-legs of any Feynmann diagrams, just like the Faddeev-Popov ghosts. Thus, it means that we can only detect the existence of these off-shell particles by measuring the radiative loop effects rather then finding these particles directly on colliders.

This idea can be generalized to other models. The new particles invented in these models can be hidden inside the loop in order to escape the detections. In general, these new particles should be assigned with charges of some unbroken symmetries in order for them to form closed loops, without any channels decaying into pure standard model (SM) particles. e.g., in Ref. \cite{MainSource}, it is the R-parity of the supersymmetric particles to play this role.

In this paper, we invent off-shell tachyons \cite{TachyonAncester, QuantizeTachyon, FermionicTachyon, FermionicTachyonQuantization} to be quantized in the unconventional way. This lead to a solution to the zero result of the loop-diagrams involving off-shell non-tachyonic particles when we apply the half-retarded and half-advanced propagators introduced in Ref. \cite{MainSource, TachyonicHalfPropagator, OtherSourceOfHalfRetardedAndHalfAdvancedPropagator}. Thus, We could not see an ``on-shell'' tachyon so that we need not worry about observing something moving faster than the light, and these off-shell tachyons contribute to the loop-diagrams, leaving us some observable effects.

\section{Ordinary Off-Shell Non-Tachyonic Particles Should Form a Closed Loop}

Without loss of generality, we introduce an unbroken $Z_2$ symmetry in this paper. Usually, the non-tachyonic $Z_2$-odd particles can decay into the lightest $Z_2$-odd particle. If some of these $Z_2$-odd particles are quantized through the unconventional way described in Ref. \cite{MainSource}, and the other $Z_2$-odd particles are quantized through the normal way, inconsistencies will be the case.

Suppose $A$ and $B$ are two $Z_2$-odd particles, and $m_A < m_B$. If both particles are quantized through the normal way, the decay $B \rightarrow A + \lbrace$CP-even particles$\rbrace$ can usually happen. The self-energy diagrams $B \rightarrow A + \lbrace$CP-even particles$\rbrace \rightarrow B$ also contain imaginary parts which contribute to the width of the $B$'s Breit-Wigner propagator $\frac{i}{p^2 - m_B^2 + i m_B \Gamma_B}$, where $\Gamma_B$ is the decay-width.

However, if $A$ is quantized through the normal way, and $B$ is quantized through the unconventional way, the diagram $B \rightarrow A + \lbrace$CP-even particles$\rbrace$ can still move the pole of the $B$'s propagator by a quantity of $i m_B \Gamma_B$, which destroys the structure of the propagator
\begin{eqnarray}
P\left(\frac{1}{p^2-m^2} \right) &=& \frac{1}{2} \left(\frac{1}{p^2-m^2 + i \epsilon} + \frac{1}{p^2-m^2 - i \epsilon}\right), \label{PrincipalValueWithoutDeltaFunction}
\end{eqnarray}
or
\begin{eqnarray}
P\left(\frac{1}{p^2-m^2} \right) &=& \frac{1}{p^2 - m^2 \pm i \epsilon} \pm i \pi \delta(p^2-m^2) \label{PrincipalValueWithDeltaFunction}
\end{eqnarray}
invented in Ref. \cite{MainSource}. These are the principal values of the propagator, and should be applied when a particle is quantized through the unconventional way.

\begin{figure}
\includegraphics[width=2in]{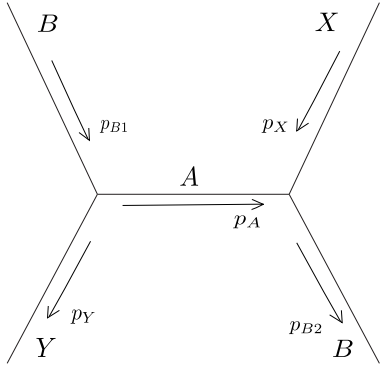}
\caption{The probable t-channel diagram in which $A$ might be near the shell. $X$ and $Y$ might be the SM-particles or other particles that do not carry the charges of the $A$ and $B$.}
\label{TChannel}
\end{figure}

If $A$ is quantized through the unconventional way, and $B$ is quantized through the normal way, the discussions are a little bit complicated. Without loss of generality, let $A$ be the lightest $Z_2$-odd particle. It should be noted that the in-lined $A$ can still be near the shell through the t-channel diagrams, e.g. $B + X \rightarrow B + Y$ in Fig. \ref{TChannel}. The $X$ and $Y$ are some $Z_2$-even particles and $m_B > m_X + m_A$, $m_B > m_Y + m_A$.

The t-channel near-shell stable particles are rarely discussed in the literature, but this case does exist. The integral over the phase space is actually infinite due to the divergence of the propagator $ \left| \frac{1}{t-m_A^2 \pm i \epsilon} \right|^2$. If $A$ and $B$ are both quantized through the normal way, this non-physical infinite can be subtracted by eliminating the on-shell $C$-effects
\begin{eqnarray}
\sigma_{\text{OS}} &=& \frac{1}{2 E_B \cdot 2 E_X |\vec{v}_B - \vec{v}_X| } \int \frac{d^3 \vec{p}_Y}{(2 \pi)^3} \frac{d^3 \vec{p}_{B2}}{(2 \pi)^3} \frac{d^3 \vec{p}_A}{(2 \pi)^3} \frac{1}{2 E_Y \cdot 2 E_{B2} \cdot 2 E_A} \nonumber \\
&& \left| \mathcal{M}_{p_{B1} \rightarrow p_Y, p_A} \right|^2 \left| \mathcal{M}_{p_{X}, p_{A} \rightarrow p_B} \right|^2 \frac{1}{2 \epsilon} (2 \pi)^4 \delta(p_{B1}-p_A-p_Y) (2 \pi)^4 \delta(p_A - p_X - p_{B2}). \label{OS_Subtraction}
\end{eqnarray}
In our appendix, we will derive (\ref{OS_Subtraction}) and will show that how the on-shell effects be subtracted. However, in our case that $A$ is quantized through the unconventional way, $A$ cannot be on-shell and thus (\ref{OS_Subtraction}) is absent, leaving us an infinite result of the diagram in Fig. \ref{TChannel}.

In a word, $A$ and $B$ should be both quantized through the normal way, or the unconventional way. In the latter case, these particles can only form a closed loop.

\section{Zero Result of the Non-Tachyonic Off-Shell Particles' Loop Diagram}

\begin{figure}
\includegraphics[width=3in]{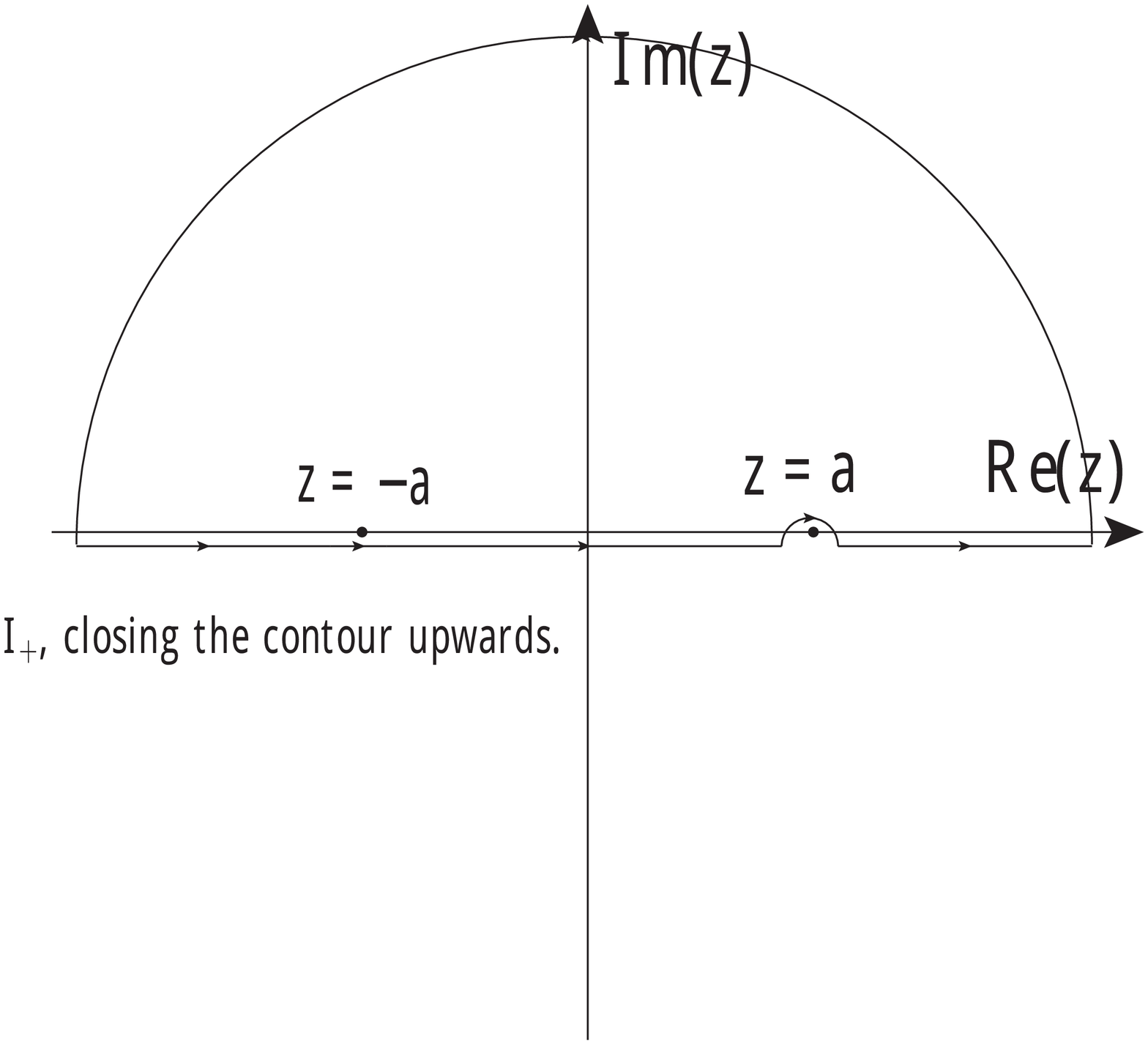}
\includegraphics[width=3in]{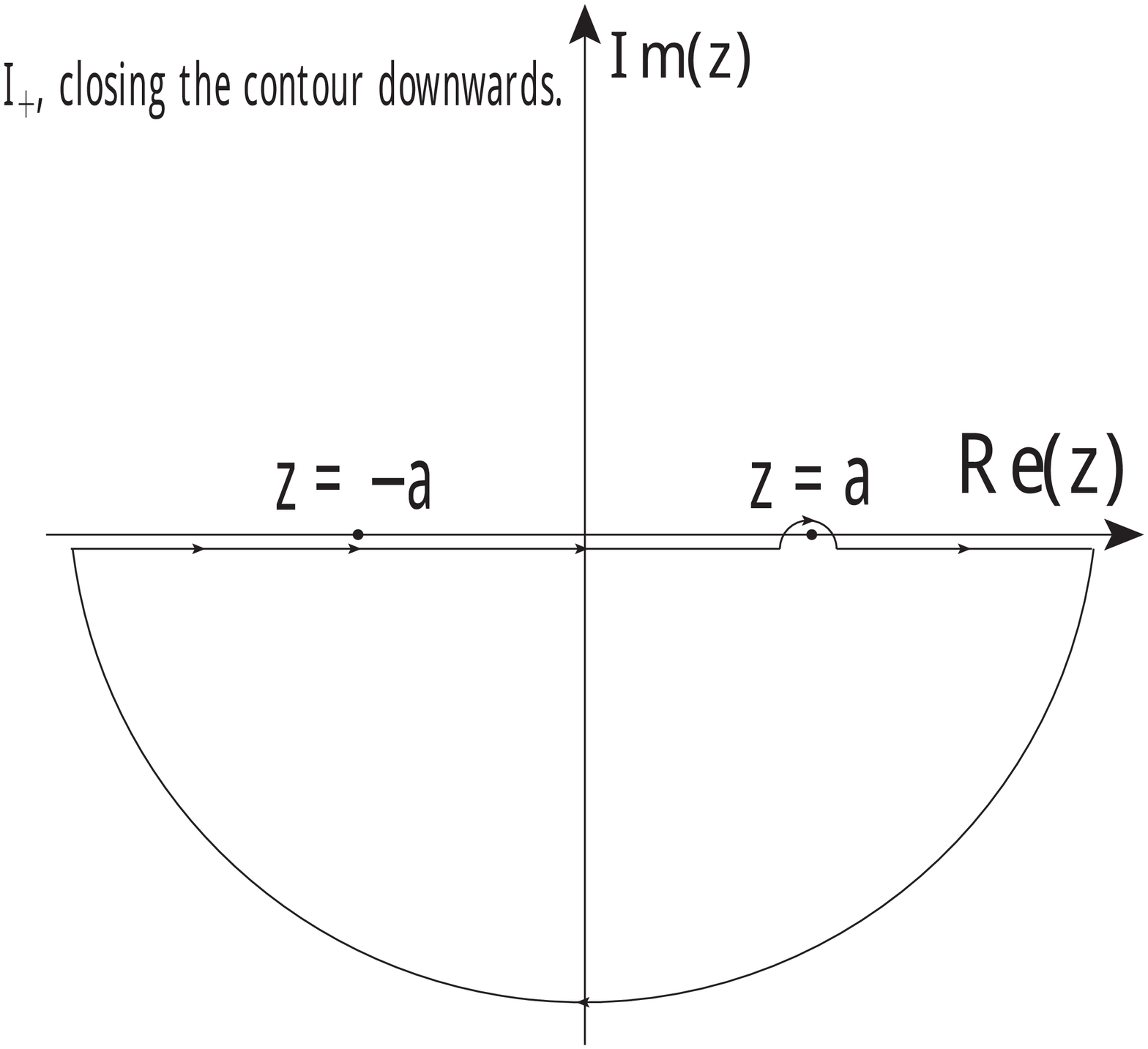}
\caption{$I_+$, different contour closing path.} \label{IPlusTraditional}
\includegraphics[width=3in]{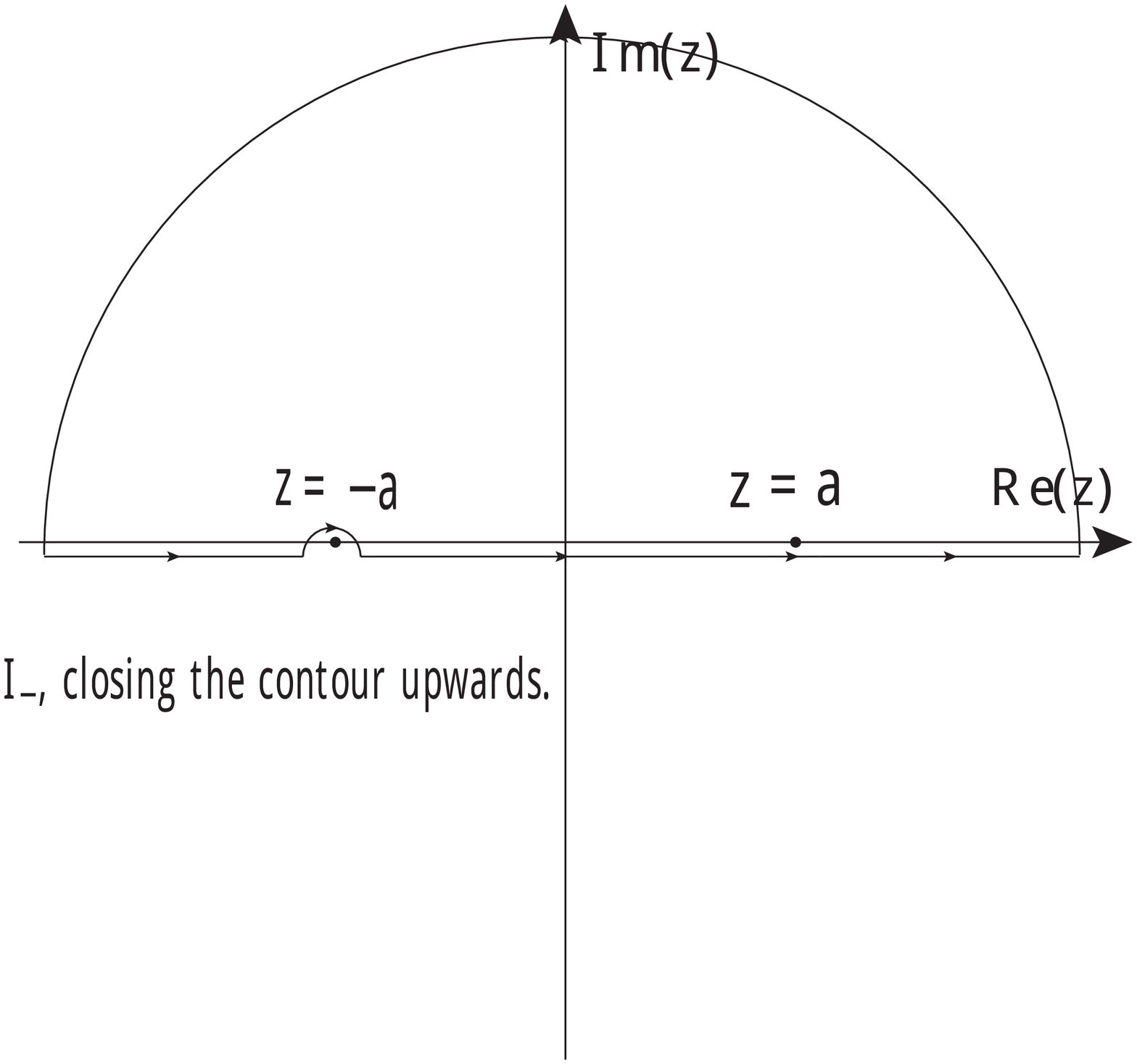}
\includegraphics[width=3in]{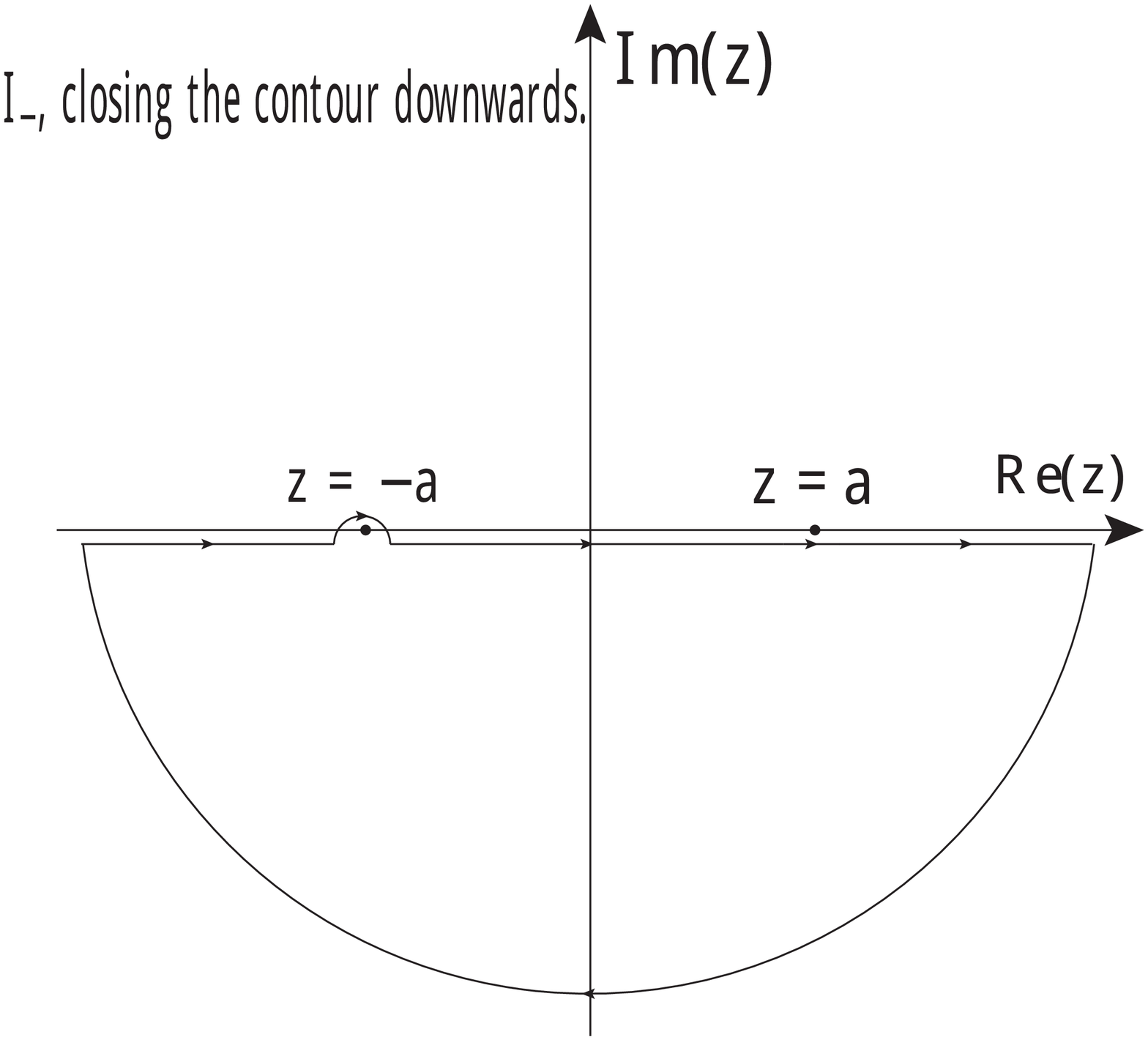}
\caption{$I_-$, different contour closing path.}\label{IMinusTraditional}
\end{figure}

Let's start from calculating this integral
\begin{eqnarray}
I_{\pm} = \int_{-\infty}^{+\infty} \frac{1}{z^2-a^2 \pm i \epsilon} dz, \label{StartIntegral}
\end{eqnarray}
where $\epsilon$ is an infinitesimal positive number introduced in order to avoid the two poles $z=\pm a$. As the integrand fades out as $\sim \frac{1}{z^2}$, one can close the contour upwards or downward to pick up the different residues as shown in Fig. \ref{IPlusTraditional}-\ref{IMinusTraditional}, resulting in the similar consequence
\begin{eqnarray}
I_{\pm} = \mp \frac{i \pi}{a}, \label{TraditionalMethod}
\end{eqnarray}
Hence,
\begin{eqnarray}
\frac{I_+ + I_-}{2} = 0.
\end{eqnarray}

Now we are going to calculate this integral in another way. Notice that $Re(I_{\pm})=0$, and the $Im(I_{\pm})$ only comes from the area near the two poles when the contour is bypassing them. Suppose there is a pole $z=a$ located on the real axis with is residue to be $Res(z=a)$, when the contour is going above this pole, it contributes a $i \pi Res(z=a)$, and when it is going beneath this pole, it becomes $-i \pi Res(z=a)$. Then for (\ref{StartIntegral}), $Res(z=-a)= -\frac{1}{2 a}$ and $Res(z=a)=\frac{1}{2 a}$, so
\begin{eqnarray}
I_{\pm} = \pm(\frac{i \pi}{2 a} + \frac{i \pi}{2 a}) = \pm \frac{i \pi}{a}, \label{AnotherMethod}
\end{eqnarray}
which is compatible with the (\ref{TraditionalMethod}).

Generalize this method to calculate
\begin{eqnarray}
I(z_1, z_2, ..., z_n, P_1, P_2, ..., P_n) = \int_{\infty}^{\infty} \frac{1}{(z-z1+i P_1 \epsilon)(z-z2+i P_2 \epsilon)...(z-z2+i P_n \epsilon)} dz ,
\end{eqnarray}
where $z_1$, $z_2$, ..., $z_n$ are real numbers which define the positions of the poles, and $P_1$, $P_2$, ...$P_n$ can be $+1$ or $-1$ which decide how the contour bypasses the poles. If $P_i=+1$, it means that the contour bypasses $z=z_i$ upwards, and if $P_i = -1$, it means that the contour bypasses $z=z_i$ downwards. Then we can immediately write down
\begin{eqnarray}
I(z_1, z_2, ..., z_n, P_1, P_2, ..., P_n) = \sum_{i=1}^{n} P_i Res(z=z_i) \pi i. \label{NonTachyonLoop}
\end{eqnarray}

\begin{figure}
\includegraphics[width=3in]{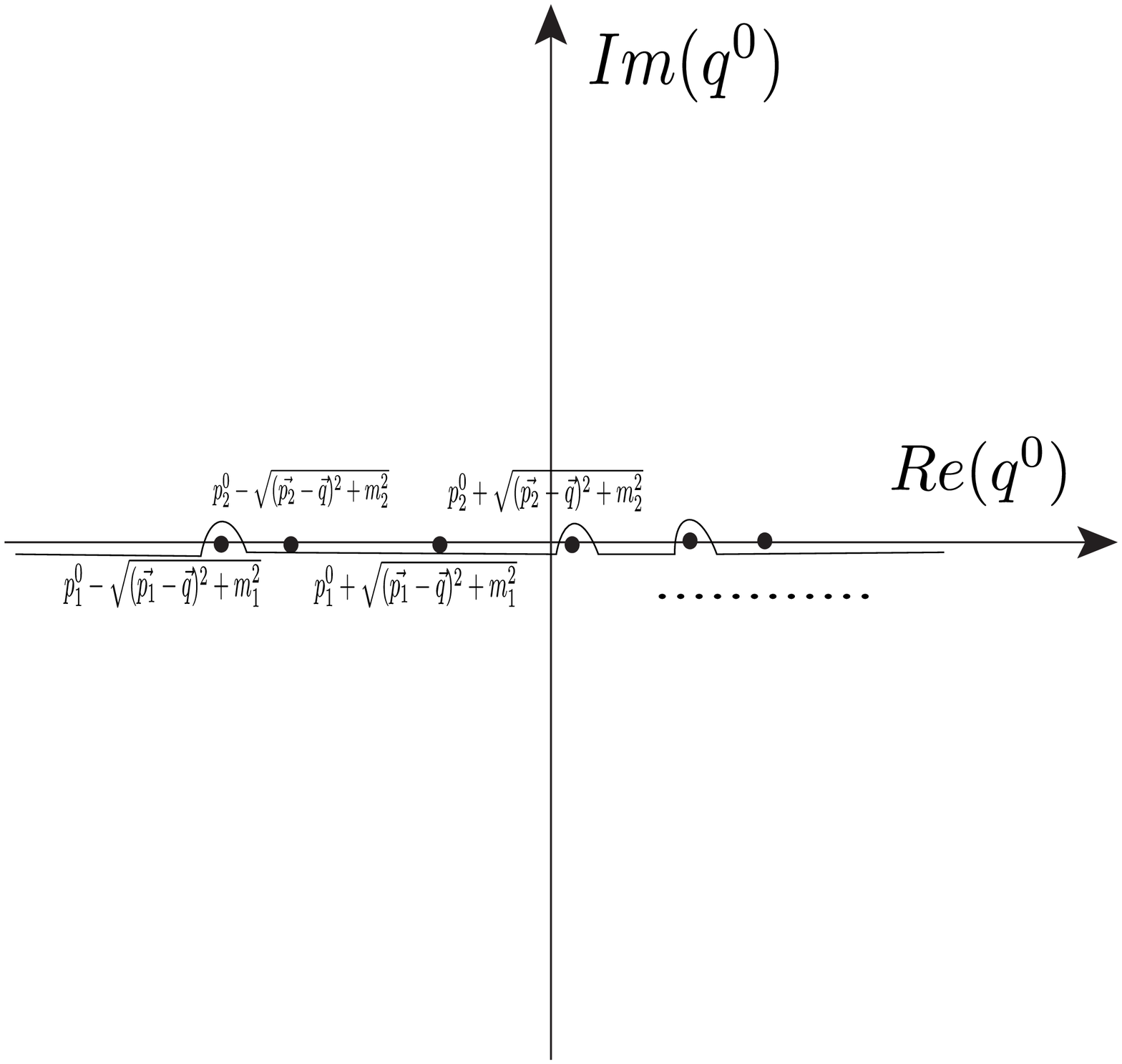}
\includegraphics[width=3in]{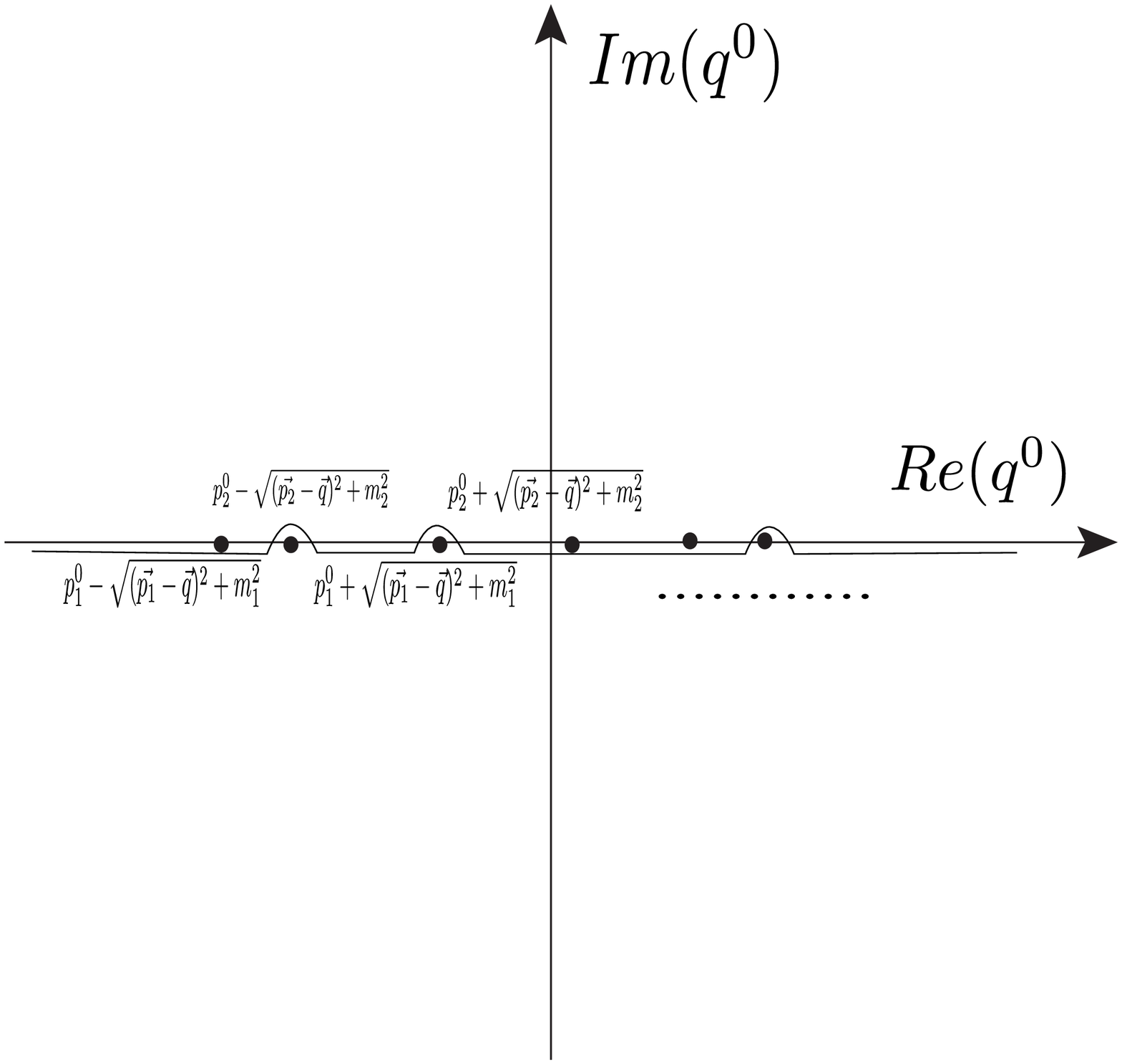}
\caption{An example of comparing the contour path of $I(P_1, P_2, ..., P_n)$ and $I(-P_1, -P_2, ..., -P_n)$. Notice that the they bypass the poles in totally opposite manners.} \label{q0IntegralOposite}
\end{figure}

Then we are prepared to calculate the loop diagrams involving the non-tachyon off-shell particles. Any of this diagrams should contain at least one subloop, each line formed by a non-tachyonic off-shell particle. To calculate this subloop, we need to calculate
\begin{eqnarray}
\mathcal{M}_{\text{sub}} = \int_{\infty}^{\infty} \frac{d^4 q}{(2 \pi)^4} P\left(\frac{i}{[(q-p_1)^2-m_1^2]}\right) P\left(\frac{i}{[(q-p_2)^2-m_2^2]}\right) ... P\left(\frac{i}{[(q-p_n)^2-m_n^2])}\right).
\end{eqnarray}
If we adopt (\ref{PrincipalValueWithoutDeltaFunction}) to calculate (\ref{NonTachyonLoop}), and integrate out $dp_0$ at first,
\begin{eqnarray}
& & \int_{-\infty}^{+\infty} \frac{dq^0}{(2 \pi)^4} P\left(\frac{i}{[(q-p_1)^2-m_1^2]}\right) P\left(\frac{i}{[(q-p_2)^2-m_2^2]}\right) ... P\left(\frac{i}{[(q-p_n)^2-m_n^2])}\right) \nonumber \\
&=& \sum_{ \lbrace P_1, P_2, ..., P_n \rbrace} I(P_1, P_2, ..., P_n ),
\end{eqnarray}
where $P_i=\pm 1$, $i=1$-$n$, and $\sum\limits_{ \lbrace P_1, P_2, ..., P_n \rbrace }$ means to enumerate all the combinations of $\lbrace P_1, P_2, ..., P_n \rbrace$ and then sum over them. The definition of $I(P_1, P_2, ..., P_n )$ is
\begin{eqnarray}
& & I(P_1, P_2, ..., P_n) \nonumber \\
&=& \int_{-\infty}^{+\infty} \frac{dq^0}{(2 \pi)^4} \frac{i^n}{2^n} \frac{1}{[(q-p_1)^2 - m_1^2 + i P_1 \epsilon] [(q-p_2)^2 - m_2^2 + i P_2 \epsilon] ... [(q-p_n)^2 - m_n^2 + i P_n \epsilon]}.
\end{eqnarray}
We are not going to talk about the massless particles, so $m_i^2 > 0$, and then $(\vec{q}-\vec{p_i})^2 + m_i^2 > 0 $ always holds. Hence, all the poles are located in the $q_0$'s real-axis. Notice that $I(P_1, P_2, ..., P_n)$ bypasses the poles in a totally opposite manner compared with $I(-P_1, -P_3, ..., -P_n)$, e.g. Fig. \ref{q0IntegralOposite}, thus
\begin{eqnarray}
I(P_1, P_2, ..., P_n) = -I(-P_1, -P_2, ..., -P_n), \label{OddPropertyOfI}
\end{eqnarray}
so
\begin{eqnarray}
& & \int_{-\infty}^{+\infty} \frac{dp^0}{(2 \pi)^4} P\left(\frac{i}{[(q-p_1)^2-m_1^2]}\right) P\left(\frac{i}{[(q-p_2)^2-m_2^2]}\right) ... P\left(\frac{i}{[(q-p_n)^2-m_n^2])}\right) \nonumber \\
&=&  \sum_{ \lbrace P_1, P_2, ..., P_n \rbrace} I(P_1, P_2, ..., P_n ) \nonumber \\
&=& \frac{1}{2} \left( \sum_{ \lbrace P_1, P_2, ..., P_n \rbrace} I(P_1, P_2, ..., P_n ) - \sum_{ \lbrace P_1, P_2, ..., P_n \rbrace} I(-P_1, -P_2, ..., -P_n) \right) \nonumber \\
&=& \frac{1}{2} \left( \sum_{ \lbrace P_1, P_2, ..., P_n \rbrace} I(P_1, P_2, ..., P_n ) - \sum_{ \lbrace -P_1, -P_2, ..., -P_n \rbrace} I(-P_1, -P_2, ..., -P_n) \right) \nonumber \\
&=& \frac{1}{2} \left( \sum_{ \lbrace P_1, P_2, ..., P_n \rbrace} I(P_1, P_2, ..., P_n ) - \sum_{ \lbrace P_1, P_2, ..., P_n \rbrace} I(P_1, P_2, ..., P_n) \right) \nonumber \\
&=& 0.
\end{eqnarray}
Thus,
\begin{eqnarray}
\mathcal{M_{\text{sub}}} \equiv 0.
\end{eqnarray}

\section{Tachyonic Cases}

(\ref{OddPropertyOfI}) holds only when $m_i^2 > 0$ and all the poles are located in the $q^0$'s real-axis. If the particles appeared in the loops are tachyons, things could be different.

Tachyons are the hypothesised particles which satisfy $E^2 = p^2 - m_{\text{Tac}}^2$, and $m_{\text{Tac}}^2>0$. In order to calculate the propagators of the off-shell tachyons, we should quantize the tachyonic fields in the unconventional way \cite{MainSource}. Unlike the normal particles, there are two different momentum areas, which are the unstable region $p^2 < m_{\text{Tac}}^2$ and the stable region $p^2 > m_{\text{Tac}}^2$. These should be treated differently.

\subsection{Scalar Tachyons}
The Lagrangian of the scalar tachyons is
\begin{eqnarray}
\mathcal{L}=\partial^{\mu} \phi \partial_{\mu} \phi + m_{\text{Tac}}^2 \phi^2
\end{eqnarray}
The unstable region $p^2 < m_{\text{Tac}}^2$ is quantized according to Ref. \cite{QuantizeTachyon}, and the stable region $p^2 > m_{\text{Tac}}^2$ is treated similar to \cite{MainSource}. Then the propagator is \cite{TachyonicHalfPropagator}
\begin{eqnarray}
\langle 0 | T\lbrace \phi(x_1) \phi(x_2) \rbrace | 0 \rangle = \int \frac{d^4 p}{(4 \pi)^4} P \left( \frac{i}{p^2 + m_{\text{Tac}}^2} \right),
\end{eqnarray}
where
\begin{eqnarray}
P \left( \frac{1}{p^2 + m_{\text{Tac}}^2} \right) = \frac{1}{p^2 + m_{\text{Tac}}^2 \pm i \epsilon} \pm i \pi \delta(p^2 + m_{\text{Tac}}^2).
\end{eqnarray}

\subsection{Spinor Tachyons}
The Lagrangian of the spinor tachyons is \cite{FermionicTachyon}
\begin{eqnarray}
\mathcal{L} = i \bar{\psi} \gamma^5 \gamma^{\mu} \partial_{\mu} \psi - m_{\text{Tac}} \bar{\psi} \psi.
\end{eqnarray}
The quantization of the tachyonic spinors is a little bit complicated. We follow Ref. \cite{FermionicTachyonQuantization},
\begin{eqnarray}
\psi(x) = \int \frac{d^3 \vec{p}}{(2 \pi)^3} \frac{1}{\sqrt{2 E_{\vec{p}}}} \sum_{\sigma} \left( a_{\vec{p}}^{\sigma} u^{\sigma} (p) e^{-i p \cdot x} + b_{\vec{p}}^{\sigma \dagger} v^{\sigma} (p) e^{i p \cdot x} \right),
\end{eqnarray}
where $\sigma = \pm 1$ is the helicity of the plane wave solutions. The $u^{\sigma}$ and $v^{\sigma}$ are normalized according to
\begin{eqnarray}
\bar{u}^{\sigma_1} (p) u^{\sigma_2} (p) &=& \sigma 2 m_{\text{Tac}} \delta^{\sigma_1 \sigma_2} \nonumber \\
\bar{v}^{\sigma_1} (p) v^{\sigma_2} (p) &=& -\sigma 2 m_{\text{Tac}} \delta^{\sigma_1 \sigma_2} \nonumber \\
u^{\sigma_1 \dagger} (p) u^{\sigma_2} (p) &=& v^{\sigma_1 \dagger} (p) v^{\sigma_2} (p) = 2 E_{\vec{p}} \delta^{\sigma_1 \sigma_2}.
\end{eqnarray}
The commutators of the operators are
\begin{eqnarray}
\lbrace a_{\vec{p}_1}^{\sigma_1}, a_{\vec{p}_2}^{\sigma_2 \dagger} \rbrace =  \lbrace b_{\vec{p}_1}^{\sigma_1}, b_{\vec{p}_2}^{\sigma_2 \dagger} \rbrace = (-\sigma) (2 \pi)^3 \delta^3 (\vec{p}_1 - \vec{p}_2).
\end{eqnarray}
The Hamiltonian operator becomes
\begin{eqnarray}
\mathcal{H} = \int \frac{d^3 \vec{p}}{(2 \pi)^3} \sum_{\sigma} \sigma \left( a_{\vec{p}}^{\sigma \dagger} a_{\vec{p}}^{\sigma} + b_{\vec{p}}^{\sigma \dagger} b_{\vec{p}}^{\sigma} \right) E_{\vec{p}}.
\end{eqnarray}
For $|\vec{p}| < m_{\text{Tac}}$ and $|\vec{p}| > m_{\text{Tac}}$, we respectively proceed the Ref. \cite{FermionicTachyonQuantization} and \cite{MainSource}, again we acquire
\begin{eqnarray}
\langle 0 | T \lbrace \psi(x_1) \bar{\psi{x_2}} \rbrace | 0 \rangle = (i \gamma^5 \gamma^{\mu} \partial_{\mu} - m_{\text{Tac}}) \int \frac{d^4 p}{(4 \pi)^4} P \left( \frac{i}{p^2 + m_{\text{Tac}}^2} \right).
\end{eqnarray}

\section{Calculation of $A_0^o$ and $B_0^o$ Functions}

All the loop diagrams can be reduced into $A_0$, $B_0$, $C_0$, ... Passarino-Veltman functions \cite{PaVe}, and only $A_0$ and $B_0$ contribute to the divergences in the usual cases, so we are going to calculate the $A_0^o$ and $B_0^o$ functions, which are the corresponding version of the $A_0$ and $B_0$ functions in the off-shell tachyonic case,
\begin{eqnarray}
A_0^o (m_{\text{Tac}}) &=& \frac{1}{i \pi^2} \int_{-\infty}^{\infty} d^4 q P \left( \frac{1}{q^2 + m_{\text{Tac}}^2} \right), \nonumber \\
B_0^o (p^2; m_{\text{Tac}1}, m_{\text{Tac}2}) &=& \frac{1}{i \pi^2} \int_{-\infty}^{\infty} d^4 q P \left( \frac{1}{q^2 + m_{\text{Tac}1}^2} \right) P \left( \frac{1}{(q+p)^2 + m_{\text{Tac}2}^2} \right).
\end{eqnarray}

\subsection{Calculation of the $A_0^o$ Function}

Let's integrate out $q^0$ at first. Notice that only in the unstable area $m^2 > \vec{q}^2$ can the pole $q_0 = \pm i \sqrt{m^2 - \vec{q}^2}$ be located on the imaginary axis, avoiding the situation of (\ref{OddPropertyOfI}).
\begin{eqnarray}
A_0^o(m_{\text{Tac}}) &=& \frac{1}{i \pi^2} \int_0^{m_{\text{Tac}}} d^3 \vec{q} \int_{-\infty}^{+\infty} \frac{dq^0}{q^{0~2}-\vec{q}^2+m^2} \nonumber \\
&=& \frac{1}{i \pi^2} \int_0^{m_{\text{Tac}}} d^3 \vec{q} \frac{1}{m^2-\vec{q}^2} \int_{-\infty}^{+\infty} \frac{dq^0}{1+\frac{q^{0~2}}{m^2-\vec{q}^2}} \nonumber \\
&=& \frac{1}{i \pi^2} \pi \int_0^{m_{\text{Tac}}} d^3 \vec{q} \frac{1}{\sqrt{m_{\text{Tac}}^2 - \vec{q}^2}} \nonumber \\
&=& \frac{1}{i \pi^2} \pi \cdot 4 \pi \int_0^{m_\text{Tac}} q^2 dq \frac{1}{\sqrt{m^2 - q^2}} = -i m_{\text{Tac}}^2 \pi.
\end{eqnarray}
We can see that there is no divergence in the result. In fact, the traditional counting of the ``divergence degree'' is applied after the Wick's rotation, which is impossible in our cases.

\subsection{Calculation of the $B_0^o$ Function}

To calculate $B_0^o$, we adopt a different form of the propagator 
\begin{eqnarray}
P \left( \frac{1}{q^2 + m_{\text{Tac}}^2} \right) = \frac{1}{q^2 + m_{\text{Tac}}^2 + i \epsilon} + i \pi \delta ( p^2 + m_{\text{Tac}}^2 ),
\end{eqnarray}
then
\begin{eqnarray}
& & B_0^o (p^2; m_{\text{Tac} 1}, m_{\text{Tac} 2}) \nonumber \\
&=&  B_0 ( p^2; i m_{\text{Tac} 1}, i m_{\text{Tac} 2} ) + B_{\delta} ( p^2; m_{\text{Tac} 1}, m_{\text{Tac} 2} ) \nonumber \\
&+& B_{\delta} ( p^2; m_{\text{Tac} 2}, m_{\text{Tac} 1} ) + B_{\delta \delta} (p^2; m_{\text{Tac} 1}, m_{\text{Tac} 2}),
\end{eqnarray}
where
\begin{eqnarray}
B_0 (p^2; i m_{\text{Tac}1}, i m_{\text{Tac}2} ) = \frac{\mu^{4-D}}{i \pi^2} \int d^D q \frac{1}{(q^2 + m_{\text{Tac}1}^2 + i \epsilon)[(q+p)^2 + m_{\text{Tac}2}^2 + i \epsilon]}
\end{eqnarray}
is the usual $B_0$ function with the traditional Feymann propagators, and
\begin{eqnarray}
 B_{\delta} ( p^2; i m_{\text{Tac} 1}, i m_{\text{Tac} 2} ) &=& \frac{\mu^{4-D}}{i \pi^2} \int d^D q \frac{1}{(q+p)^2 + m_{\text{Tac} 2}^2 + i \epsilon} i \pi \delta ( q^2 + m_{\text{Tac} 1}^2 ), \nonumber \\
B_{\delta \delta} (p^2; m_{\text{Tac} 1}, m_{\text{Tac} 2}) &=& \frac{1}{i \pi^2} \int d^4 q i \pi \delta(q^2 + m_{\text{Tac} 1}^2) i \pi \delta( (q+p)^2 + m_{\text{Tac} 2}^2 ).
\end{eqnarray}

To calculate the $ B_0 ( p^2; i m_{\text{Tac} 1}, i m_{\text{Tac} 2} )$, the traditional tricks involving Feynmann integral and Wick's rotation are applied,
\begin{eqnarray}
& & B_0 ( p^2; i m_{\text{Tac} 1}, i m_{\text{Tac} 2} ) \nonumber \\
&=& \frac{2}{\varepsilon} - \gamma + \ln 4 \pi + \int_0^1 dx ln \left( \frac{\mu^2}{p^2 (x^2-x) - m_{\text{Tac}1}^2 x - m_{\text{Tac}2}^2 (1-x) - i \epsilon }\right), \label{UsualB0}
\end{eqnarray}
where $\varepsilon = 4-D$.

$B_{\delta \delta} (p^2; m_{\text{Tac} 1}, m_{\text{Tac} 2})$ is just cancelling all the imaginary part of $B_0 ( p^2; i m_{\text{Tac} 1}, i m_{\text{Tac} 2} )$ according to the optical theorem.

To calculate $B_{\delta} ( p^2; m_{\text{Tac} 2}, m_{\text{Tac} 1} )$, we work in the $p = (p0, \vec{0})$ reference frame, then
\begin{eqnarray}
& & B_{\delta} ( p^2; m_{\text{Tac} 2}, m_{\text{Tac} 1} ) \nonumber \\
&=& \frac{\mu^{4-D}}{\pi} \frac{2 (\pi)^{\frac{D-1}{2}} }{\Gamma \left( \frac{D-1}{2} \right)} \int_{m_{\text{Tac}1}}^{+\infty} q^{D-2} dq \left[ \frac{1}{(2 E_{\vec{q}})(p_0^2 + 2 E_{\vec{q}} p_0 - m_{\text{Tac}1}^2 + m_{\text{Tac}2}^2 + i \epsilon) } \right. \nonumber \\
& & + \left. \frac{1}{(2 E_{\vec{q}}) (p_0^2 - 2 E_{\vec{q}} p_0 - m_{\text{Tac}1}^2 + m_{\text{Tac}2}^2 + i \epsilon)} \right]
\nonumber \\
&=& \frac{\mu^{4-D}}{\pi} \frac{2 (\pi)^{\frac{D-1}{2}} }{\Gamma \left( \frac{D-1}{2} \right)} \int_{0}^{+\infty} \frac{(E_{\vec{q}}^2 + m_{\text{Tac}1}^2)^{\frac{D-3}{2}}}{2} dE_{\vec{q}} \left[ \frac{1}{p_0^2 + 2 E_{\vec{q}} p_0 - m_{\text{Tac}1}^2 + m_{\text{Tac}2}^2 + i \epsilon } \right. \nonumber \\
& & + \left. \frac{1}{p_0^2 - 2 E_{\vec{q}} p_0 - m_{\text{Tac}1}^2 + m_{\text{Tac}2}^2 + i \epsilon} \right],
\nonumber \label{BDeltaInt}\\
\end{eqnarray} 
where $E_{\vec{q}} = \sqrt{\vec{q}^2 - m_{\text{Tac}1}^2}$.

The full calculations of (\ref{BDeltaInt}) are too complicated to be discussed in this paper. We only note that in the complex-$E_{\vec{q}}$ plane, the integral can be divided into $\int_0^{+\infty} = \int_0^{\pm i m_{\text{Tac}1}} + \int_{\pm i m_{\text{Tac}1}}^{\infty}$. The $\int_0^{\pm i m_{\text{Tac}1}}$ part is finite and the $\int_{\pm i m_{\text{Tac}1}}^{\infty}$ part contributes to
\begin{eqnarray}
B_{\delta} ( p^2; m_{\text{Tac} 2}, m_{\text{Tac} 1} )_{\text{div}} = -\frac{2 \left( m_{\text{Tac}2}^2 - m_{\text{Tac}1}^2 + p_0^2 \right)}{2 p_0^2 \varepsilon},
\end{eqnarray}
which means
\begin{eqnarray}
B_{\delta} ( p^2; m_{\text{Tac} 2}, m_{\text{Tac} 1} )_{\text{div}} + B_{\delta} ( p^2; m_{\text{Tac} 1}, m_{\text{Tac} 2} )_{\text{div}} = -\frac{2}{\varepsilon},
\end{eqnarray}
that cancels the divergent part of (\ref{UsualB0}) accurately.

\section{Summary}
We have proved that the loop contributions from the half-retarded and half-advanced propagators of the off-shell particles are always zero unless these particles are tachyons. We have calculated the Passarino-Veltman $A_0^o$ and $B_0^o$ functions of these particles and showed that all the divergent parts have been cancelled. The loop effects of the probably existing off-shell tachyonic particles are non-zero and thus might be detected in the future.

\begin{acknowledgements}

We would like to thank Professor Chun Liu, Professor Yi Liao, Professor Jian-Ping Ma, Professor Deshan Yang, Dr.~Jia-Shu Lu for helpful discussions.
This work was supported in part by the National Natural Science Foundation of China under Nos. 11375248, and by the
National Basic Research Program of China under Grant No. 2010CB833000.

\end{acknowledgements}

\appendix

\section{the Calculations of the t-channel Diagrams with a On-shell Normally Quantized Mediator}

Now we are going to calculate the cross-section of the diagram in Fig. \ref{TChannel},
\begin{eqnarray}
\sigma_t &=& \frac{1}{2 E_B \cdot 2 E_X | \vec{v}_B - \vec{v}_X |} \int \frac{d^3 \vec{p}_{B2}}{(2 \pi)^3} \frac{d^3 \vec{p}_{Y}}{(2 \pi)^3} \frac{1}{2 E_{\vec{p}_{B2}} \cdot 2 E_{\vec{p}_{Y}}}| F_{ABY} F_{ABX} |^2 \nonumber \\
& &\frac{i}{p_A^2 - m_A^2 + i \epsilon} \frac{-i}{p_A^2 - m_A^2 - i \epsilon} (2 \pi)^4 \delta^4 (p_{B1} + p_X - p_Y - p_{B2}), \label{Pure_T_Channel}
\end{eqnarray}
where $F$'s are the coupling constants. Here we are not going to talk about CP violation effects so these coupling constants are assigned with real numbers without loss of generality. Insert $1=\int \frac{d^4 p_A}{(2 \pi)^4} (2 \pi)^4 \delta^4 (p_A + p_Y - p_{B1})$ into (\ref{Pure_T_Channel}).
\begin{eqnarray}
\sigma_t &=& \frac{1}{2 E_B \cdot 2 E_X | \vec{v}_B - \vec{v}_X |} \int \frac{d^3 \vec{p}_{B2}}{(2 \pi)^3} \frac{d^3 \vec{p}_{Y}}{(2 \pi)^3} \frac{d^3 \vec{p}_A}{(2 \pi)^3} | F_{ABY} F_{ABX} |^2 \frac{1}{2 E_{\vec{p}_{B2}} \cdot 2 E_{\vec{p}_{Y}}} \nonumber \\
& &\int \frac{d p_A^0}{(2 \pi)} (2 \pi)^4 \delta^4 (p_A + p_X - p_{B2}) \frac{i}{p_A^2 - m_A^2 + i \epsilon} \frac{-i}{p_A^2 - m_A^2 - i \epsilon} (2 \pi)^4 \delta^4 (p_{B1} - p_{Y} - p_{A}).
\end{eqnarray}
When we are trying to integrate out $d p_A^0$, the contour passes by the poles $p_A^0 = \sqrt{\vec{p_A}^2 + m_A^2 + i \epsilon}$ and $p_A^0 = -\sqrt{\vec{p_A}^2 + m_A^2 - i \epsilon}$. Pick up the contributions from these two poles and then we acquire the divergent part of the $\sigma_t$,
\begin{eqnarray}
\sigma_{\text{OS}} &=& \frac{1}{2 E_B \cdot 2 E_X |\vec{v}_B - \vec{v}_X| } \int \frac{d^3 \vec{p}_Y}{(2 \pi)^3} \frac{d^3 \vec{p}_{B2}}{(2 \pi)^3} \frac{d^3 \vec{p}_A}{(2 \pi)^3} \frac{1}{2 E_Y \cdot 2 E_{B2} \cdot 2 E_A} \nonumber \\
&& \left| \mathcal{M}_{p_{B1} \rightarrow p_Y, p_A} \right|^2 \left| \mathcal{M}_{p_{X}, p_{A} \rightarrow p_B} \right|^2 \frac{1}{2 \epsilon} (2 \pi)^4 \delta(p_{B1}-p_A-p_Y) (2 \pi)^4 \delta(p_A - p_X - p_{B2}), \label{OS_Subtraction_Appendix}
\end{eqnarray}
which is just the (\ref{OS_Subtraction}).

\begin{figure}
\includegraphics[width=1.5in]{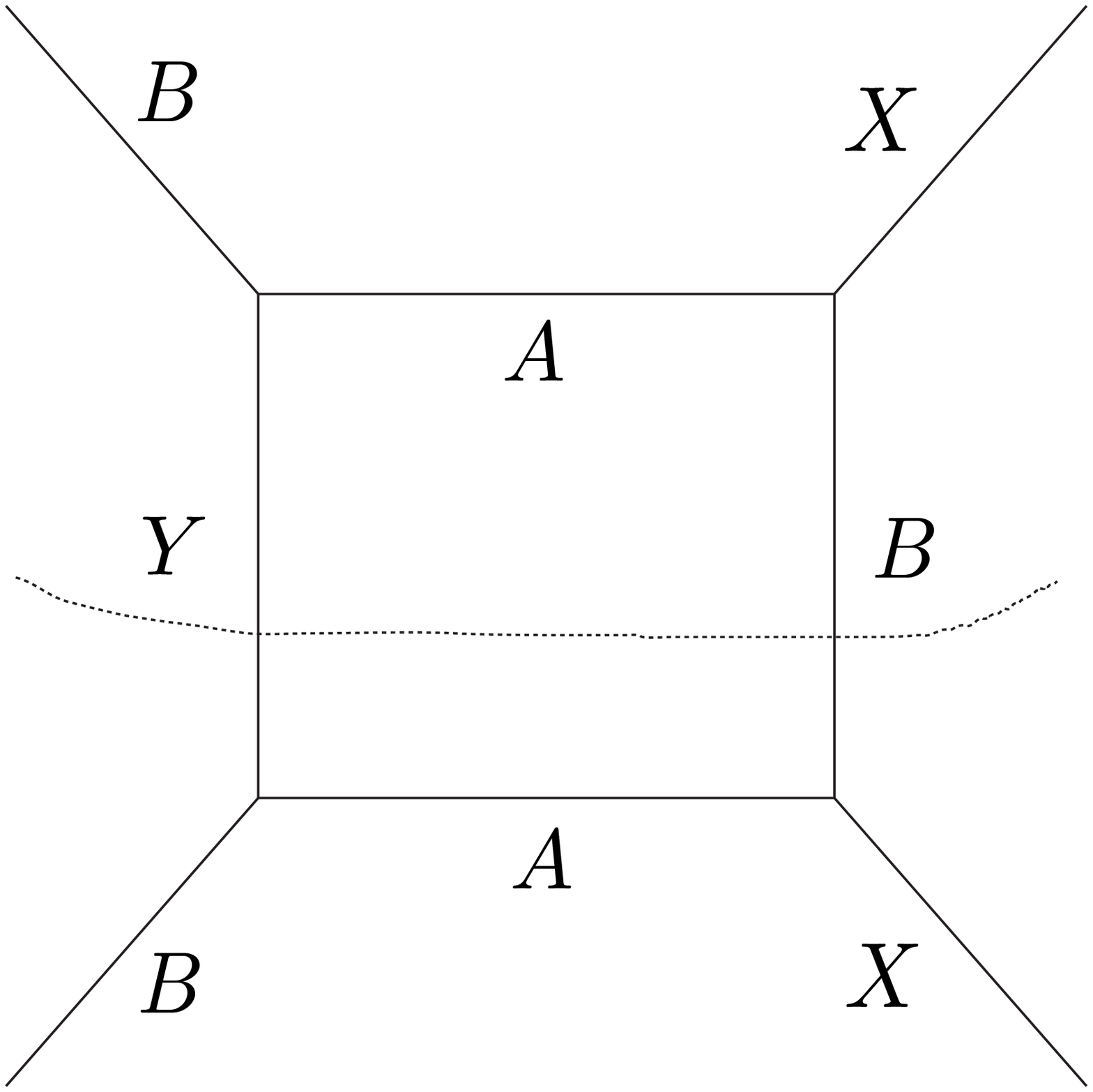}
\caption{One way to cut the box diagram. This relates to the t-channel in Fig. \ref{TChannel}. \label{BoxT}}
\includegraphics[width=1.5in]{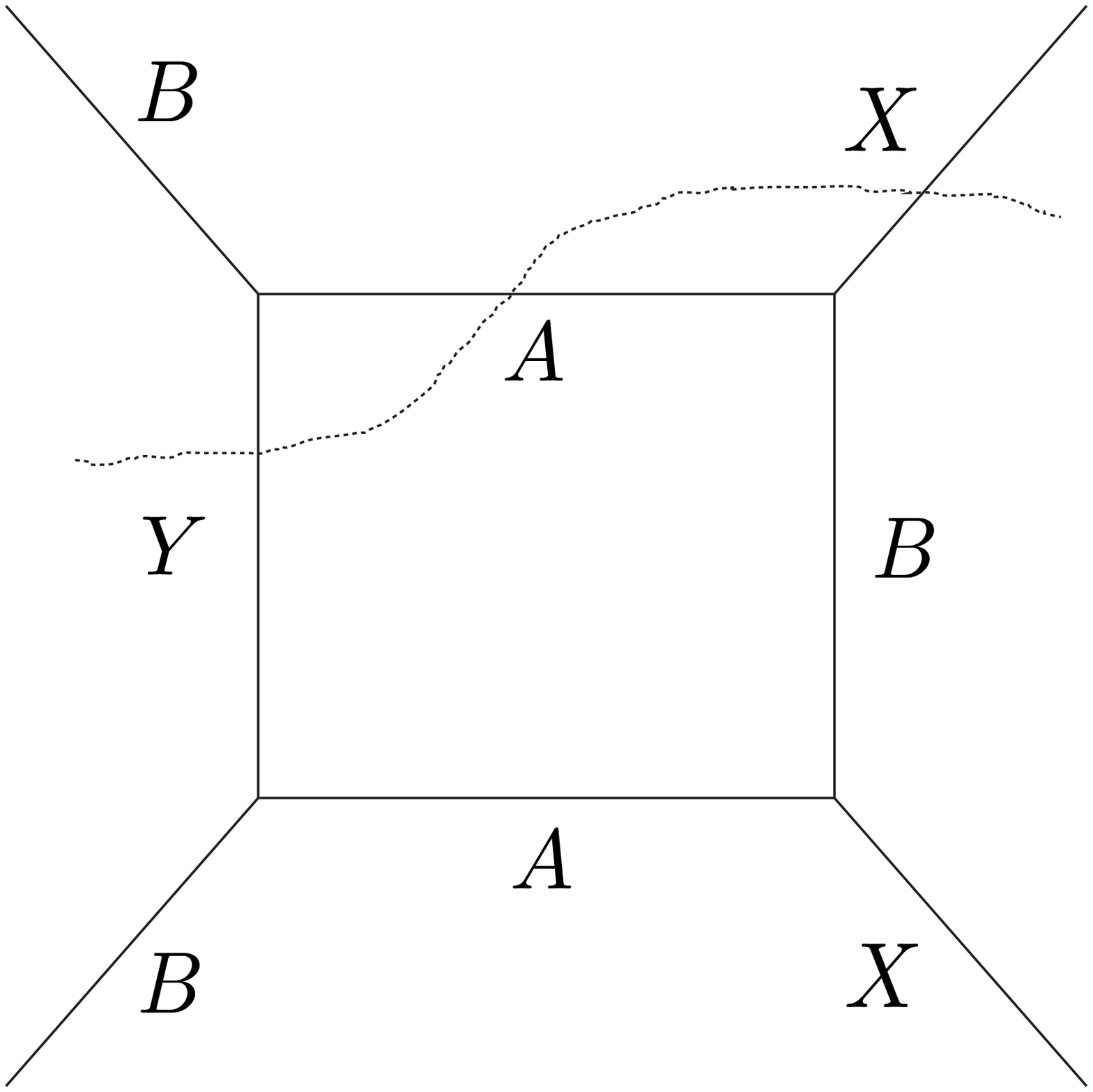}
\includegraphics[width=1.5in]{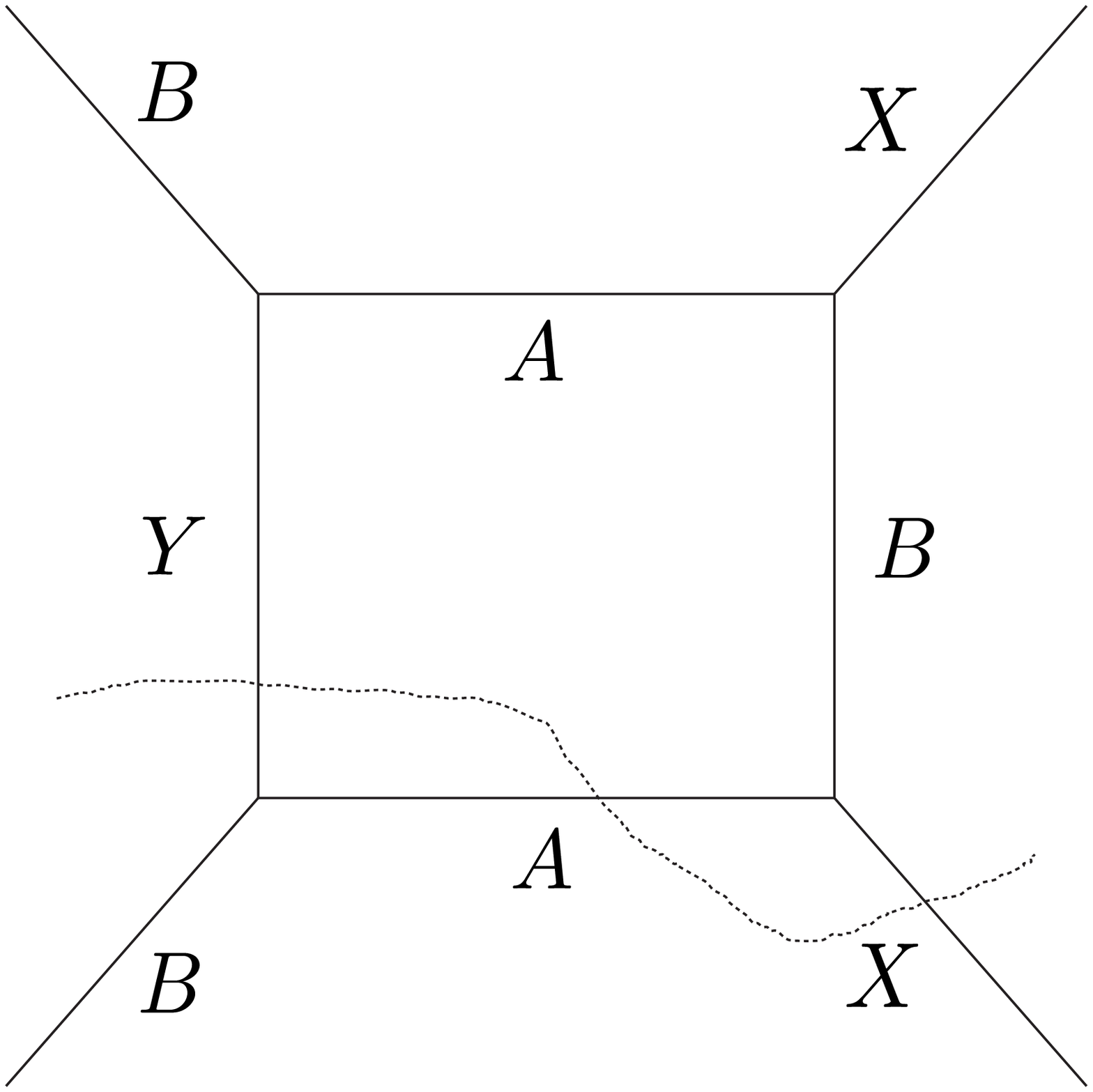}
\includegraphics[width=1.5in]{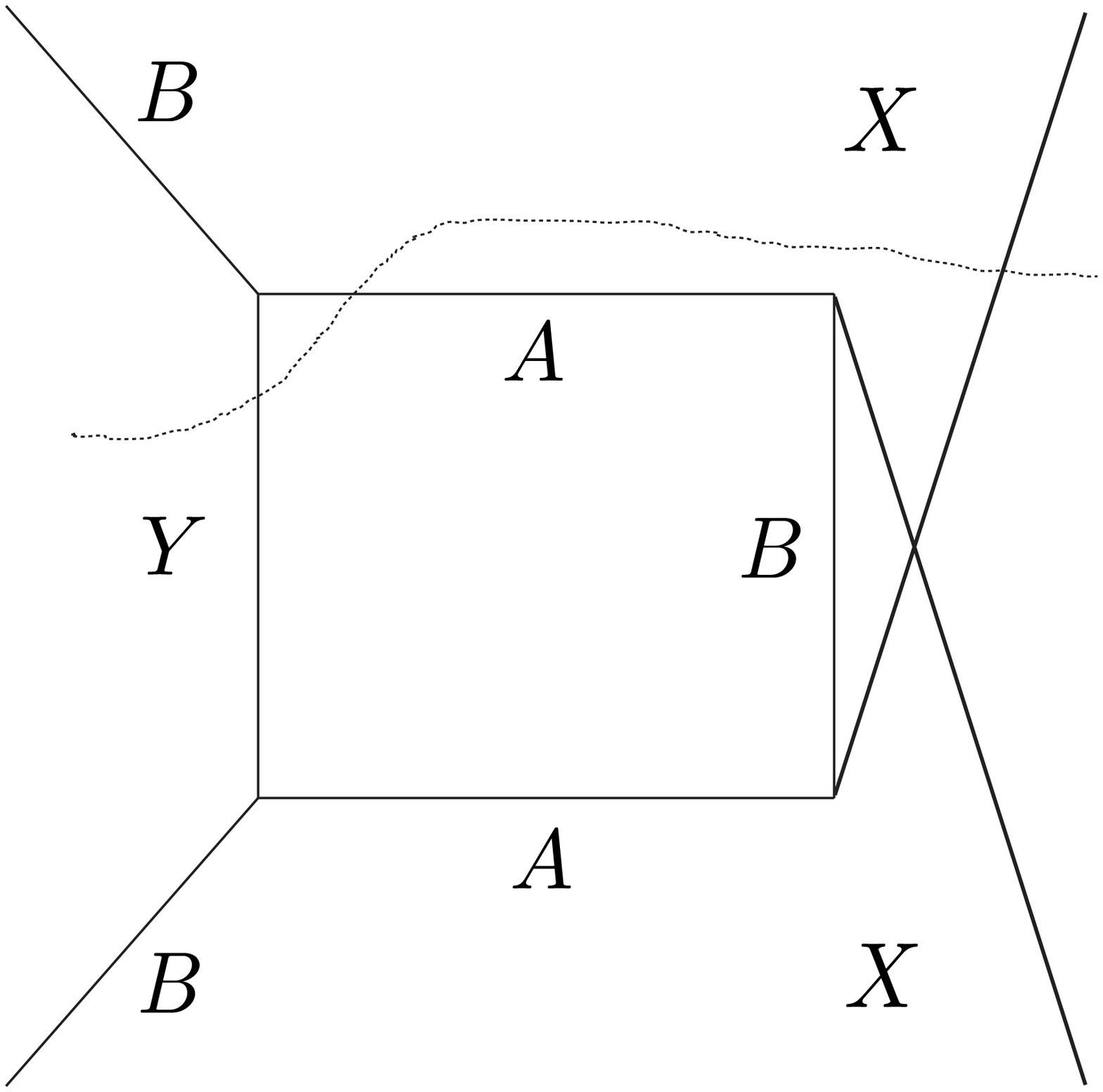}
\includegraphics[width=1.5in]{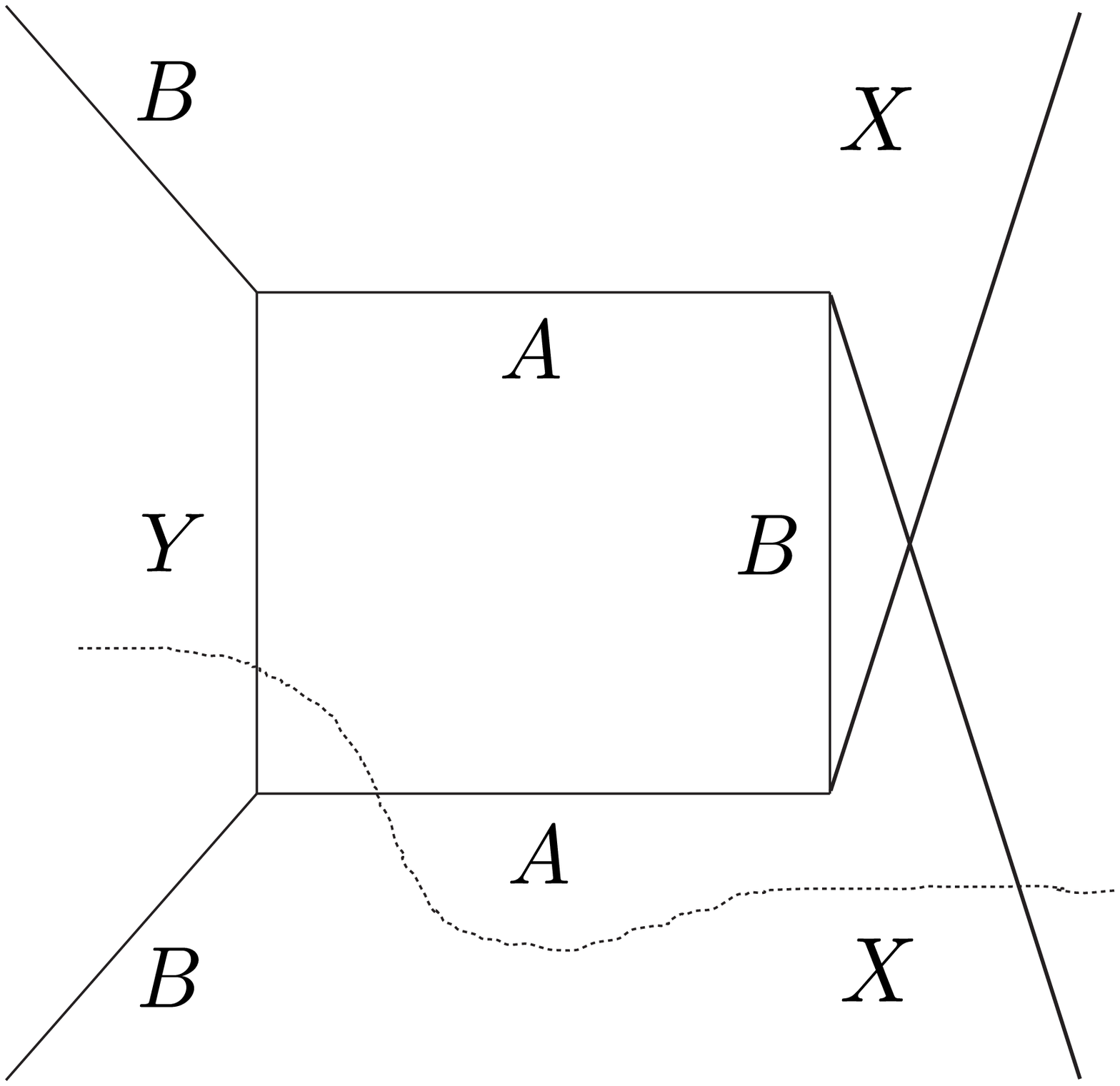}
\caption{Other ways to cut the box diagram. \label{OtherBox}}
\end{figure}

\begin{figure}
\includegraphics[width=1.5in]{tchannel.eps}
\includegraphics[width=1.5in]{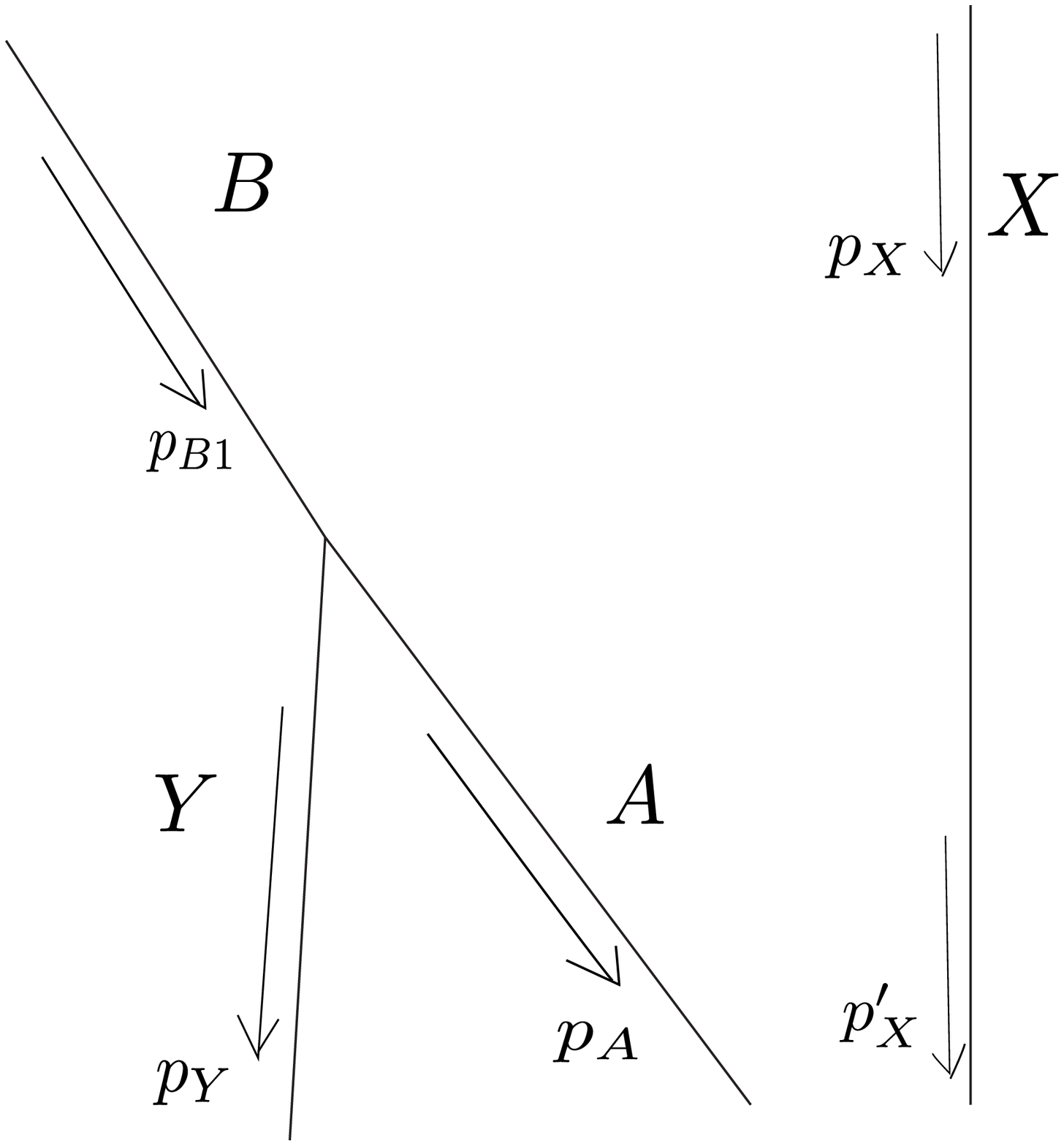}
\includegraphics[width=1.5in]{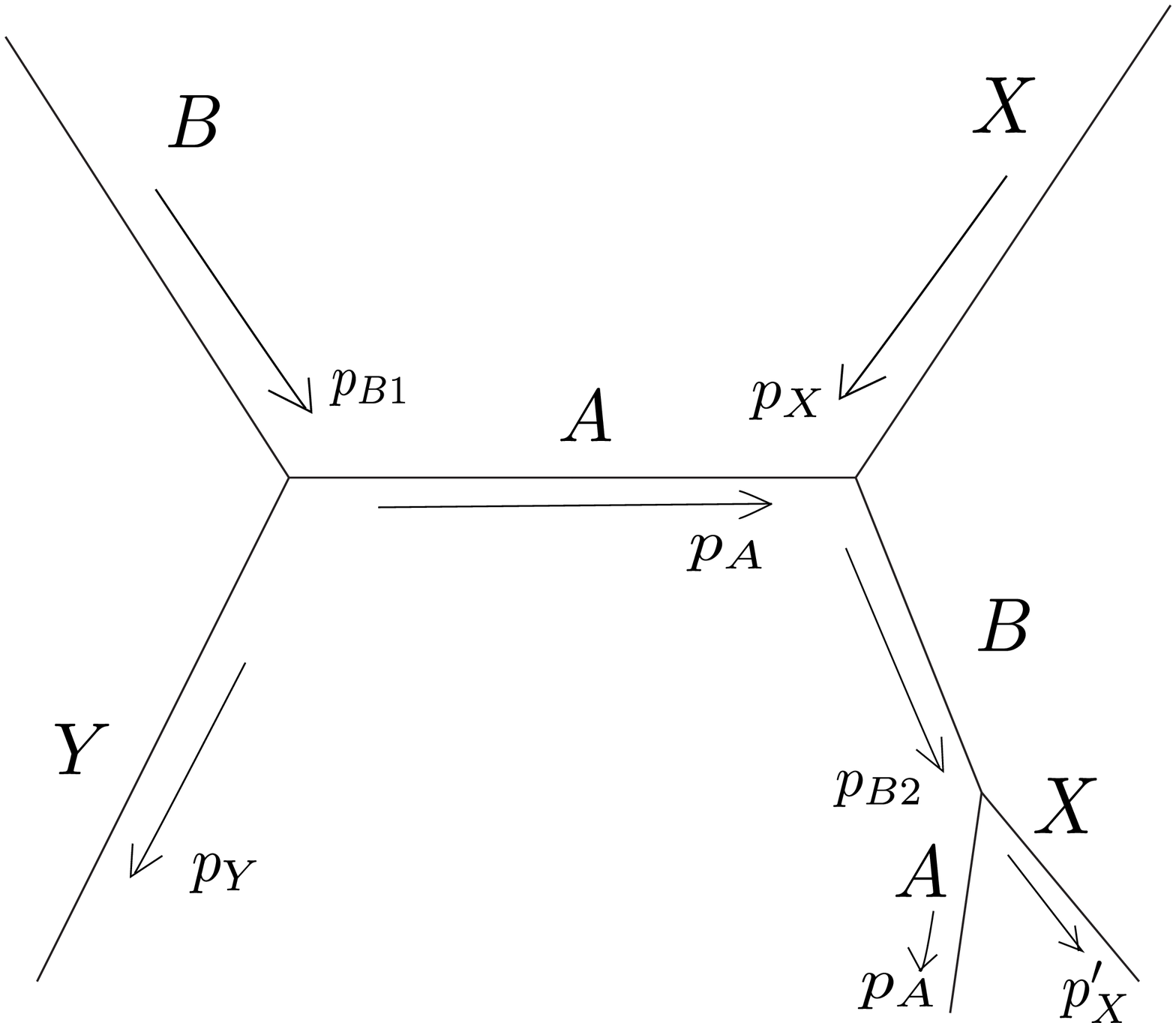}
\includegraphics[width=1.5in]{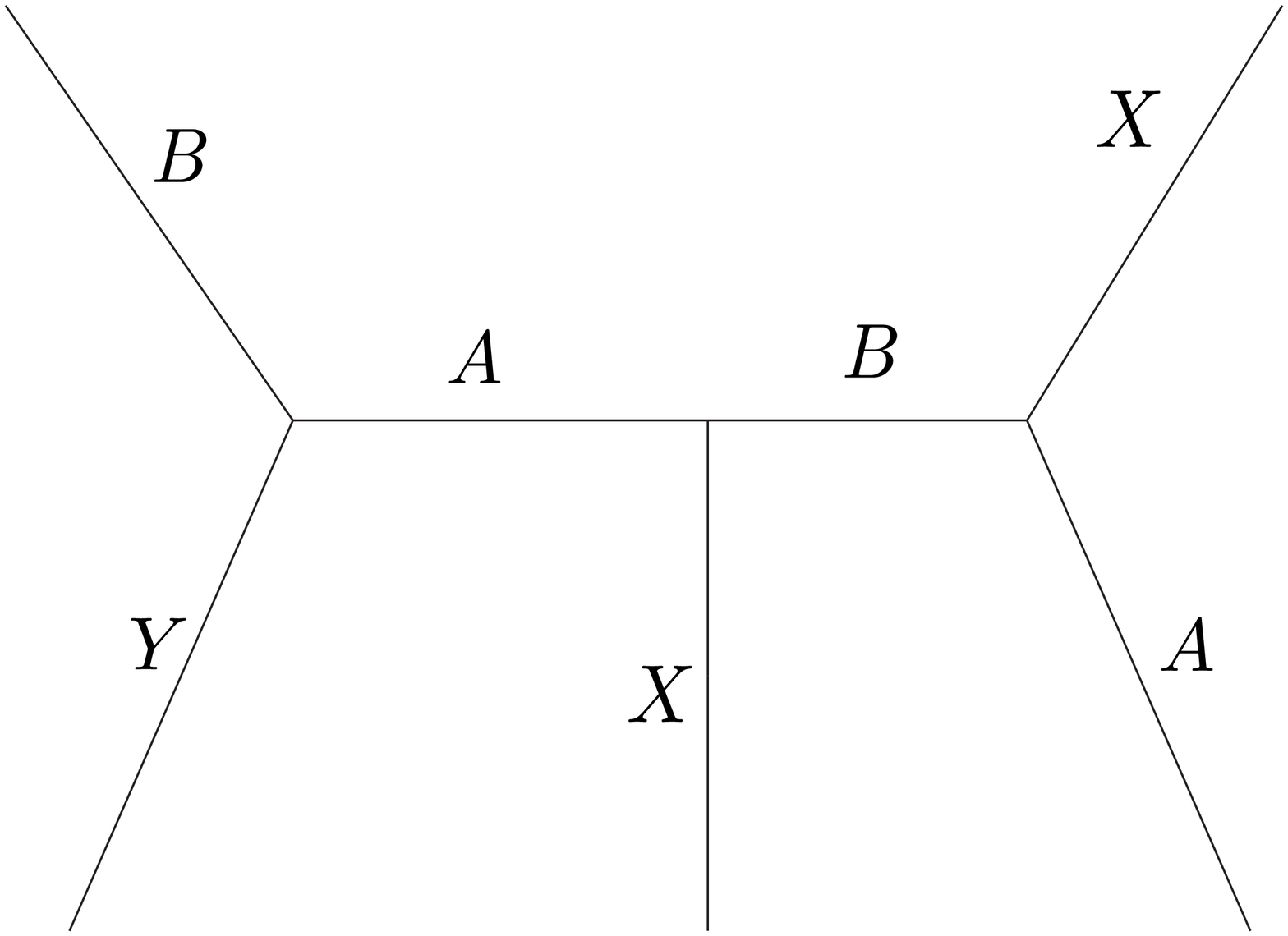}
\caption{Tree-level diagrams contributed to the total cross-section. \label{TreeLevelDiagrams}}
\end{figure}

This divergence is due to the on-shell effects of $A$, and is related to one way to cut off the box diagram in Fig. \ref{BoxT}. Besides this, there exist other box diagrams and other ways to cut them \cite{Cutkosky}. See Fig. \ref{OtherBox}. It means that we should sum over these diagrams in Fig. \ref{TreeLevelDiagrams}. Hence, some interference terms occur and appear to cancel the (\ref{OS_Subtraction_Appendix}).

The final-state particles in these diagrams are different. However, $B$ may decay as it is taking part in the process, so we need to consider the complete styles of the final-state particles, making it necessary to sum over all the diagrams in Fig. \ref{TreeLevelDiagrams}. 

The middle diagram in Fig. \ref{TreeLevelDiagrams} is weird. It is not only disconnected but also contains a ``bare'' line, the $X$. It means that this $X$ does not take part in the scattering process and by applying the Lorentz-invariant inner products of the one-particle states, we acquire the ``Feynmann-rules'' of this ``bare'' line,
\begin{eqnarray}
\left\langle p_{X} | p_{X}^{\prime} \right\rangle = 2 E_{\vec{p}_{X}} (2 \pi)^3 \delta^3 (\vec{p}_{X} - \vec{p^{\prime}}_{X}).
\end{eqnarray}

We are now prepared to calculate the interference term $ \left( \text{\includegraphics[width=0.6in]{DecayMove.eps}} \right) \cdot \left( \text{ \includegraphics[width=0.6in]{TwoToThree.eps}} \right)^*$,
\begin{eqnarray}
\sigma_{\text{interfere}} &=& \frac{1}{2 E_B \cdot 2 E_X | \vec{v}_B - \vec{v}_X |} \int \frac{d^3 \vec{p}_Y}{(2 \pi)^3} \frac{d^3 \vec{p}_A}{(2 \pi)^3} \frac{d^3 \vec{p^{\prime}}_X}{(2 \pi)^3} \frac{1}{2 E_{\vec{p}_Y} \cdot 2 E_{\vec{p}_A} \cdot 2 E_{\vec{p}_X}} \nonumber \\
& & i F_{ABY} \cdot (-i F_{ABY}) \cdot (i F_{ABX}) \cdot (i F_{ABX}) \frac{-i}{p_A^2 - m_A^2 - i \epsilon} \frac{-i}{p_{B2}^2 - m_{B}^2 - i \epsilon} \nonumber \\
& &(2 \pi)^3 \delta^3 (\vec{p}_{X} - \vec{p^{\prime}}_{X}) (2 \pi)^4 \delta^4 ( p_{B1} - p_{Y} - p_{A} ) \nonumber \\
&=&\frac{1}{2 E_B \cdot 2 E_X | \vec{v}_B - \vec{v}_X |} \int \frac{d^3 \vec{p}_Y}{(2 \pi)^3} \frac{d^3 \vec{p}_A}{(2 \pi)^3} \frac{1}{2 E_{\vec{p}_Y} \cdot 2 E_{\vec{p}_A}} i F_{ABY} \cdot (-i F_{ABY}) \nonumber \\
& &\cdot (i F_{ABX}) \cdot (i F_{ABX}) \frac{-i}{p_A^2 - m_A^2 - i \epsilon} \frac{-i}{p_{B2}^2 - m_{B}^2 - i \epsilon} (2 \pi)^4 \delta^4 ( p_{B1} - p_{Y} - p_{A} ).
\end{eqnarray}
Again we are going to insert $1=\int \frac{d^4 p_{B2}}{(2 \pi)^4} (2 \pi)^4 \delta^4 (p_X + p_A - p_{B2})$ into the integral,
\begin{eqnarray}
\sigma_{\text{interfere}} &=& \frac{1}{2 E_B \cdot 2 E_X | \vec{v}_B - \vec{v}_X |} \int \frac{d^3 \vec{p}_Y}{(2 \pi)^3} \frac{d^3 \vec{p}_A}{(2 \pi)^3} \frac{d^3 \vec{p}_{B2}}{(2 \pi)^3} \frac{1}{2 E_{\vec{p}_Y} \cdot 2 E_{\vec{p}_A}} i F_{ABY} \nonumber \\
& &\cdot (-i F_{ABY}) \cdot (i F_{ABX}) \cdot (i F_{ABX}) \int \frac{dp_{B2}^0}{2 \pi} \frac{-i}{p_A^2 - m_A^2 - i \epsilon} \frac{-i}{p_{B2}^2 - m_{B}^2 - i \epsilon} \nonumber \\
& & (2 \pi)^4 \delta^4 (p_X + p_A - p_{B2}) (2 \pi)^4 \delta^4 (p_X + p_A - p_{B2}),
\end{eqnarray}
and then integrate out the $d p_{B2}^0$. Notice that the divergent part is contributed from near the pole $p_{B2}^0 = \sqrt{ \vec{p}_{B2}^2 + m_B^2 + i \epsilon}$. Pick up the residue there, and notice the $\delta^4 (p_X + p_A - p_{B2})$ keeps $p_A$ a little bit away from the mass-shell, we acquire
\begin{eqnarray}
\sigma_{\text{interfere}} |_{\text{div}} = - \frac{1}{2} \sigma_{\text{OS}},
\end{eqnarray}
which means $ \left( \text{\includegraphics[width=0.6in]{DecayMove.eps}} \right) \cdot \left( \text{ \includegraphics[width=0.6in]{TwoToThree.eps}} \right)^* + \left( \text{\includegraphics[width=0.6in]{DecayMove.eps}} \right)^* \cdot \left( \text{ \includegraphics[width=0.6in]{TwoToThree.eps}} \right)$ accurately cancels the divergent term of $\sigma_t$.

Following the similar process, we can also prove that the divergent parts of the rest of the interference terms $ \left( \text{\includegraphics[width=0.6in]{DecayMove.eps}} \right) \cdot \left( \text{ \includegraphics[width=0.6in]{TwoToThreeOther.eps}} \right)^* + \left( \text{\includegraphics[width=0.6in]{DecayMove.eps}} \right)^* \cdot \left( \text{ \includegraphics[width=0.6in]{TwoToThreeOther.eps}} \right)$ equals zero. We omit the detailed proof in this paper because this appendix is aimed at proving the cancellation of the t-channel on-shell divergences in Fig. \ref{TChannel}.

From the discussions above, we can clearly learn that it is the B's on-shell decay effects that play the role of cancelling the divergence of the t-channel on-shell divergences. 


\begin{thebibliography}{99}
\bibitem{ClaimToBeEarlest}
  The idea that the superpartners can only appear in the loop was mentioned in
  D.~M.~Ghilencea,
  Nucl.\ Phys.\ B {\bf 876}, 16 (2013)
  [arXiv:1302.5262 [hep-ph]].
\end{thebibliography}
\end{document}